# Complete Security Framework for Wireless Sensor Networks

Kalpana Sharma, M.K. Ghose, Kuldeep

Sikkim Manipal Institute of Technology
Majitar-737136, Sikkim, India,
kalpanasharma.smit@gmail.com, hod_cse@smu.edu.in

*Abstract*— **Security concern for a Sensor Networks and level of security desired may differ according to application specific needs where the sensor networks are deployed. Till now, most of the security solutions proposed for sensor networks are layer wise i.e a particular solution is applicable to single layer itself. So, to integrate them all is a new research challenge. In this paper we took up the challenge and have proposed an integrated comprehensive security framework that will provide security services for all services of sensor network. We have added one extra component i.e. Intelligent Security Agent (ISA) to assess level of security and cross layer interactions. This framework has many components like Intrusion Detection System, Trust Framework, Key Management scheme and Link layer communication protocol. We have also tested it on three different application scenarios in Castalia and Omnet++ simulator.**

## I. INTRODUCTION

Wireless Sensor Networks are being employed in various real time fields like Military, disaster management, Industry, Environmental Monitoring and Agriculture Farming etc. Due to diversity of so many real time scenarios, security for WSNs becomes a complex issue. For each implementation, there are different type of attacks possible and demands a different security level. Major challenge for employing an efficient security scheme comes from the resource constrained nature of WSNs like size of sensors, Memory, Processing Power, Battery Power etc. and easy accessibility of wireless channels by good citizens and attackers.

Although research in the sensor network security area is progressing at tremendous pace [15]; still there is lack of an integrated comprehensive framework which can provide security services to each layer and services of sensor networks. Current research in this area majorly focuses on providing layered solutions, which can provide security service for one layer only. Also some solutions address particular kind of attacks only.

In a diverse application field of sensor networks, specifically application designer knows which data needs to be secured with which kind of security service [12].We can take example of two popular WSN applications like Agriculture Farming and Military Surveillance system whereas in case of agriculture farming only data integration (HASH functions) check can do, but military surveillance needs security services like encryption, authentication and strong resilience to node compromise attacks. By all means, a security setup for an application must always be subject to a thorough security evaluation in order to justify its security promises and to foster the application developer's awareness regarding which aspects are secure and which are at risk, thus avoiding a false sense of security. For a reasonable security evaluation, we have added another logical component in sensor node structure namely ISA (Intelligent Security Agent) which will asses security level needs of a particular sensor network deployment.

In this paper, section 2 describes current approaches in security of sensor networks and their limitations, section 3 formulates the security framework problem and its design goals, section 4 introduces all the component of framework and section 5 describes the simulation results and analysis of ISA. Finally section 6 concludes with future work.

## II. RELATED WORKS

Extensive research is being carried out to address security issues in sensor networks like link layer communication protocol, Intrusion Detection Scheme, Secure Routing protocol, Trust Models, freshness transmission, key management schemes etc. In this section a brief overview of the security solutions available in the literature are summarized. One common approach to create secure platforms in WSN is by providing link layer cryptographic primitives or libraries. TinySec [17], Secure Sense, and MiniSec[16] are examples of this approach. Although MiniSec provides energy efficient security compared to other link layer solutions the main drawback of MiniSec is in terms of providing same security level to each application scenarios, thus lacking adaptive security or scenario specific security. A very good technique for low overhead freshness transmission using bloom filter and last bit optimization is given in MiniSec.

TinySec and Secure Sense assume to have *a global common secret key* among the nodes which is assigned before the deployment of the network and is used to provide security services such as encryption and authentication in link layer. The main drawback with this approach is that it is not resistant against node capture attacks in which an adversary can pollute an entire sensor network by compromising only one single node. In SenSec, there are three types of keys: *Global Key*, *Cluster Key* and *Sensor key*. The global key is generated by the base station, pre-deployed on each sensor node and shared by all nodes. This key is used to broadcast messages in the network. However, this protocol again falls prey to node capture attacks in which dedicated attackers can find this global key and broadcast commands or data to the network. The provision of maximum level of security for all types of communication in each sensor node, as the one which appears in TinySec and MiniSec, is not suitable for use as in a general security platform for WSN since it can lead to unnecessary waste of system resources and noticeably reduces the network lifetime. Although there has been an attempt made in Secure Sense to address this issue; its solution cannot be well integrated with higher level services appropriately. In other solutions, like secure information routing protocols such as SPINS [19] and LEAP [25] or security-aware middleware services such as secure localization or secure time synchronization [8] cryptographic key management plays an important role. Generally there are three major approaches for key management in WSN namely: Deterministic pre assignment, Random pre-distribution and Deterministic post-deployment derivation.

Examples of the first approach are SPINS [19] and LEAP in which unique symmetric keys shared by the nodes with the base station are assigned before the network is deployed. Using this approach, cryptographically strong keys can be generated; however, this involves a significant pre-deployment overhead and is not scalable.

Random-key distribution schemes like those in, PIKE [20] refer to probabilistically establishing pair wise keys between neighboring nodes in the network. However in this approach, a node has to store large number of keys. Bhaskaran Raman et al [26] pointed out that WSN protocols are very deeply dependent on Application scenarios, but most of protocols does not cite or use any specific application in its design. So current security schemes also lacks in providing security to specific scenarios while assessing their security needs. There are some approaches which addresses only routing problem like secure spin, secure sensor network routing and some other geographic techniques. Tae Kyung Kim et. al [27] gives a simple trust model using fuzzy logic that can effectively address the secure routing problem. It calculates the evaluation value for each path and ensures that packet is always forwarded to a high evaluation value path. A scheme for preventing compromised node to become cluster head is proposed by Garth et.al, which is based on trust factor. Some initiatives to provide security framework, which integrates two or more security schemes like secure cluster formation [23], key management [18] and secure routing [22], also combines link layer secure communication protocol with key distribution scheme. Security platform proposed in [22] provides defense against node compromised attacks, but does not give any mechanism to isolate them. It supports holistic security approach to provide security to WSNs. But major disadvantage of holistic security approach is that it tries to implement security layer wise which results in redundant security.

An example to fully understand the concept of redundant security is to consider a black hole attack. The security mechanisms provided in this case is both in network as well as link layer. Without a systematic view such approaches would provide redundant security thus wasting resources and unintentionally launch a SSDoS (Security Service DoS) attack. Also when data is processed layer by layer, there can be different security provisions layer wise, thus providing redundant security. Although much progress has been made for the past few years, the field remains fragmented, with contributions dispersed over seemingly disjoint yet actually connected areas,

Currently much of work is going on providing layered security for such as the Holistic Security Approach [2]. A holistic approach aims at improving performance, security, longevity with respect to changing environmental condition with some basic principles. For example in a given network security is to be ensured for all the layers of the protocol stack as shown in fig1 and also the cost of security should not be more than assessed security risks. But major disadvantage with holistic security is that it is layered and tries to implement security mechanisms for each layer, which results in wastage of power, memory, processing power and introduce message delay.

### III. PROBLEM FORMULATION

Generally a security platform that copes with constrained resources of nodes while being flexible and lightweight eases the application development process and contributes to widespread deployment of sensor networks. In order to provide such a platform we have made a few reasonable assumptions. We also suppose that the base station is safe and adversaries cannot compromise it. Our approach does not place *any* trust assumption on the communication apart from the obvious fact that there is a nonzero probability of delivering messages to related destinations. We introduce the following design goals for a practical security framework in sensor networks.

Robust, Simple and flexible Designs: - Security design should build trustworthy system out of untrustworthy components and should have ability to detect and function when need arises. Design should have minimum software bugs. Security framework should also work if we add new nodes in the network thus providing scalability.

*Component Based Security***:-**Some kind of security measures must be provided to all the components of a system as well as to network .We should concentrate on securing the whole chain.

*Adaptive Security***:-**WSNs are having numerous combination of sensing, communication and computing technologies and sensors are deployed from very sparse to dense. So depending on traffic characteristics and environment they have to adapt themselves. For ex.:- In a good environment where probability of security attacks is low, we should use low level of security. In other words, we can say that sensor node should adapt them according to outside environment. However we further categorize the notion of adaptive security into following terms.

a) Application based:- As already described in previous section that each application requires different level of security like Military Surveillance, Habitat Monitoring etc.

b) Data Based:- Level of security also depends on the type of data like there should be different level of encryption for

routing, sensed data, control packet data and encryption key information.

c) *QoS with Security*: - One important question is how to trade off between QoS parameters while providing security. Unfortunately, existing security designs can address only a small, fixed threshold number of compromised nodes; the security protection completely breaks down when the threshold is exceeded.

d) *Realistic Design***:-** Current Security design lacks this design requirement because they have an explicit threat model in mind. We have to do real trace analysis for all kind of practical attacks possible for a particular real time scenario**.**

## IV. SOLUTION MODEL

We have already seen the flaws that can be occurred in implementing layered security approach. So in this paper we would rather concentrate on cross layer security framework. In some of the recent works, there are cross layer implementation for power management schemes, path redundancy based security [21], Energy equivalence routing and various key management schemes. In support of cross layer security approach let us concentrate on the following points [3].

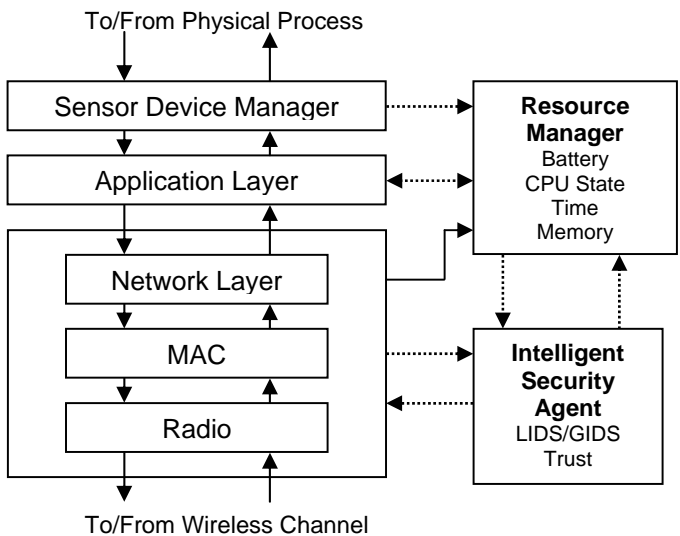

Figure 1 Node structure with ISA incorporated

1. If we want that routing should be energy efficient then we have to take care of routing in network layer, minimization of the number of control packets and retransmission in link layer and putting energy transceivers On/Off in physical layer.

2. Key management schemes make sure that all the communicating nodes possess required keys for encrypted communication. At the same time to make sure that packet reaches destination, a secure link with multi path routing is required

There are various other reasons for adopting cross layer security approach and to name some we've heterogeneous requirements and services of applications domain, cross layer intrusion detection, detection of selfish nodes and non redundant security. However, Cross layer security introduces a significant overhead in maintaining interfaces between various protocol layers for exchanging parameters. However this overhead will be much lesser in comparison with strict layered architectures. To further reduce the overhead created by cross layered architectures, we've introduced ISA (Intelligent Security Agent) to follow the recommendations given in Section 3 and 4 and to provide energy efficient and non redundant security operation while keeping protocol layer abstraction intact. ISA will be used as a separate component in node architecture (See fig 1), which can exchange parameters with all protocol layers like a Resource Manager.

We know that in a component based security framework, security is to be ensured for all the components and services in a system. So we will also address the following requirements of WSN Security.

- Robust Trust Framework using Cross Layer Approach.
- Trust Based Group Head Election.
- Key Management Architecture.
- Adaptive Secure Communication Protocol.
- Intrusion Detection System.

In addition to the above mentioned points the following assumptions are made for the proposed Trust-Framework.

- We assume TDM (Time Division Multiplexing) scheduling for communication within a group. In TDM, in a particular interval, a node will transmit otherwise it will listen passively in promiscuous mode. So a node can hear neighborhood transmission/reception.

- Each node in a network is identified by a set of Group id (8 Bits) and Node id (8 Bits) i.e. {Groupid, Nodeid}.So node communication is limited to group only.

- Each node has three different types of keys viz. a Node Based Keys which are used to listen to broadcast made by Group Head, Pair Wise Keys which are used to facilitate communication between pair wise nodes and Broadcast Keys which are the Keys used for broadcasting.

### A. Trust Framework and Group Head Election

In our proposed security scheme a network is divided into various groups and each group has its group head. Normally, to maximize network life time, a node with highest energy is chosen to be group head. Because a group head has to perform several other operations like data aggregation etc, the rate of consumption of power is very high in case of group head. We apply rotating group head so that when a node falls short of energy, it will transfer its responsibility to some other node in the group by election or some other measures. Here the security concern which arises is as follows; let us take consider the parameter 'available energy' as a measure to transfer group head responsibility. A compromised node or an adversary node would always show higher amount of energy.So there is a high probability to select an adversary as a group head. Most of the current clustering techniques assume that all the nodes are trustworthy in a network. Hence, we should choose a technique in which probability of selection of compromised nodes as group head is very low. In such a technique, a node will continuously monitor its neighbors and maintains a parameter table. All the parameters values are collected from cross layer interactions. Depending on table parameters, it will compute a trust level of all its neighbors. Table parameters are given as

TABLE I.  TRUST PARAMETERS USED IN TRUST FRAMEWORK

| Sl. No | Parameters | Node1 | Node2 | Node m |
|---|---|---|---|---|
| 1. | Available Energy (AE) | | | |
| 2. | Packet Signal Strength (PSS) | | | |
| 3. | Control Packet Received for forward(CRF) | | | |
| 4. | Control Packet Received forwarded(CRAF) | | | |
| 5. | Routing Cost(RC) | | | |
| 6. | No. of Packet Collision (NPC) | | | |
| 7. | Data Packet Received for forward (DRF) | | | |
| 8. | Data Packet Received forwarded (DRAF) | | | |
| 9. | Packet Dropped (PD) | | | |
| 10. | No. of Packets Transmitted (NPT) | | | |
| 11. | No. of Packets Received (NPR) | | | |

- $CRF_i$ - Control Packet Received for forward for a particular node i. where i=1, 2… m. same notations apply to CRAF, DRF, DRAF, NPT, and NPR.
- $AE_i$ (T1) – Available Energy for a node i at a time T1. Same notations also apply for PSS and RC.

Now the trust values from these parameters are calculated as follows:

$A_1 = (AE_i (T1) - AE_i (T2)) / AE_i (T1)$      where T1 < T2.

$A_2 = (PSS_i (T1) - PSS_i (T2)) / PSS_i (T1)$      where T1 < T2.

$A_3 = CRAF_i / CRF_i$

$A_4 = DRAF_i / DRF_i$

$A_5 = 1 - NPC_i / NPT_i$

$A_6 = 1 - PD_i / NPR_i$

Here $T_i$, the Trust Level of Node, is calculated by the node which is maintaining above table.

$T_i = w1* A_1 + w2* A_2 + w3*A_3 + w4* A_4 + w5* A_5 + w6*A_6$

Here w1, w2, w3, w4, w5, w6 are constants, whose value is chosen such that $T_i < 1$.

More on Trust based cluster scheme can be obtained from [23]. For our work we considered some extra parameters as well as discarded some of them while calculating trust value to make our scheme more generalized and robust. After computing trust level of each neighbor, a node will use these values for routing purposes also. Steps for choosing group head are as follows

- Step-1: Whenever a group head finds that it is unable to bear the group head responsibility due to some reasons like low energy etc. then it will broadcast a message for a re-election.
- Step-2: If a node gets election message, then it will find a neighbor with highest trust value and sends this to current group head as a vote.
- Step-3: Now, group head will assign this responsibility to the node having the highest number of votes. For greater integrity, a vice group head can also be chosen but not necessarily. A vice group is needed because sometimes there can be failure of new elected group head before transferring responsibility. Ids of group head and vice group head will directly be broadcasted to all group members using a secret key. All communication described above must be done by using appropriate keys as described in the next section.

*B. Key Management Architecture*

The scheme proposed by Hamed et. al [14] provides a strong defense to node compromised attacks, while being very simple to implement. But the major drawback of this scheme is that it does not provide any mechanism for changing keys periodically, because it derives all the three types of keys from the key given before deployment. There is also no mechanism to isolate compromised node from a network. So we propose a modified key management protocol which keeps all the advantages provided by Hamed et. al and at the same time try to remove some of the drawbacks of [14]. We assume that K is a key that all sensor nodes initially have. At the time of initial deployment following algorithm is executed:

**Algorithm Key Management (K: Master Key)**

// Master key is that a node has from deployment time.

Begin

A node i will broadcast its id encrypted by key K to all the neighbors.

Suppose a node j is neighbor of node i, then in its response it will also sent its id encrypted by key K.

Node i will compute all of its keys

$NB_i = F (i \| Group Head ID \| K)$,

$PW_{i,j} = F (min(i,j) \| max(i,j) \| K)$,

$BC_i = F (i \| K)$.  Here $\|$ is concatenation operator.

In the same way node j and all other neighbors will calculate the above three types of keys using master key K.

Similarly all other nodes will calculate required keys, after which they'll have to delete the master key to be resilient against node capture attacks.

End

In key establishment protocol by Hemed et. al, they send $PW_{i,j}$, $BC_i$ encrypted using $NB_j$. But to reduce communication overhead, we choose to compute them at j, because communication is three orders of magnitude more expensive than computation. Now if the same keys are to be used all the times in a network and a node becomes compromised, although the effect will be limited only to neighborhood, it can send group head a false information or launch a DoS (Denial of Service) attack to decrease the energy level of its neighbors. So we should isolate compromised nodes as soon as possible. To isolate them we propose following hierarchical Key Revocation Algorithm.

At the starting of each session, Base Station will distribute a key to the corresponding cluster heads; Cluster head will generate a new key K1= F (Group ID || Base Station Address || K) and subsequently multicasts the key K1.But the key K1 will be sent only to those nodes having high trust value. So in the previous session, if a node becomes compromised then its trust value will decrease automatically and it will not get hold of new key.

## C. Adaptive Secure Communication Protocol

None of the link layer security protocol proposed till now provides adaptive security. So we have proposed an adaptive security protocol, which dynamically adjusts itself to a particular security level depending on the network state. Mechanism of providing adaptive security is handled by the proposed component ISA (Intelligent Security Agent), which is used to make cross layer interactions easier.

*Packet Format*

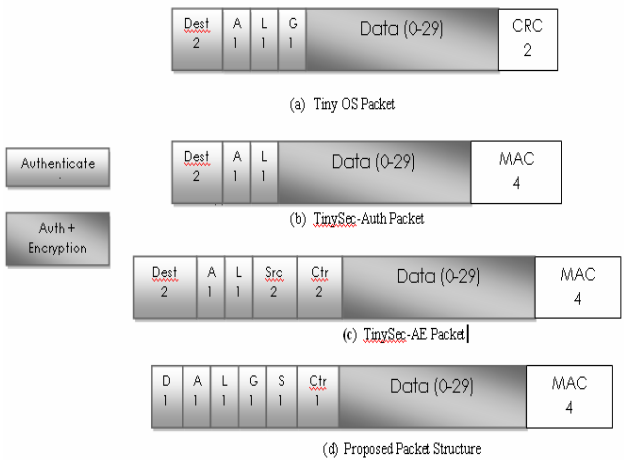

Figure -2 Comparative Analysis of Packet Structure of Different Link Layer Protocols

We have already described the limitations of current link layer protocols in Section 2. We will describe our protocol based on desired security properties that a protocol should possess.

Tinysec[17] uses CBC (Cipher Block Chaining) mode to provide message authentication (CBC-MAC).It minimizes the cryptographic primitives, but Tinysec-AE has to perform two CBC mode encryption and CBC mode authentication at a sender side and a CBC-MAC mode authentication and CBC mode decryption at receiver side, which requires two symmetric key operations cycles to compute encryption and MAC. This overhead can be reduced greatly by using authenticated –encryption method such as OCB, CBC-X etc. We use OCB to generate cipher text as well as MAC in only one symmetric key operation (Also used in Minisec).

In Tinysec[17] packets, Source and Destination field is of 2 bytes, so we can say that a network can support $2^{16}$=65536 number of nodes. We have introduced Group field because use of group field is crucial for many applications in sensor networks and also if we use a group field of 8 bits then a network can support 256 different groups, and using source id and destination id of 8 bits, a group can support 256 different nodes.

In our scheme, the number of nodes that a network can support is 256*256=65536 nodes, same as that of Tinysec. Packet overhead due to source and destination id in Tinysec is of 2+2=4 bytes. But in our scheme it is 3 bytes (including group filed).It is because a node can be distinguished by using {Groupid , Sourceid} or {Groupid, Destinationid}.This is consistent with our previous assumption that the communication of a node is limited to its group members only. Maximal payload length in Tinysec packet can be of 29bytes.So payload length cannot be greater than 5 bits and MSB 3bits are unused in each data packet to be sent. We have utilized MSB 3bits for providing adaptive security. First 2 bits will be used for encryption level and the third bit will be used for authentication purpose.

Now, we will discuss how the communication protocol mentioned above preserves the required security properties.

TABLE 2: DIFFERENT LEVEL OF ENCRYPTION

| Bits Rep | Level | Operation |
|----------|---------|------------|
| 00 | Level -0 | Simple XOR |
| 01 | Level-1 | RC5/80/4 |
| 10 | Level-2 | RC5/80/8 |
| 11 | Level-3 | RC5/80/12 |

Here RC5/80/4 represents RC5 encryption algorithm with key size 80 bits and encryption rounds 4.

*Data Secrecy and Authentication*: - We are using RC5 as block cipher for encryption coupled with OCB, so that both encryption and authentication are achieved in only one pass, thus saving in processing time and energy at sensor node. Sensor node uses encryption to provide data secrecy, but sometimes there is very little difference in consecutive readings of a sensor node .We have used a **nonce** as a counter, which is used in encryption, thus ensuring each time different cipher text is generated. We have used four level of encryption to provide adaptive security. Level of encryption will be provided by ISA.

*Replay Protection and Freshness Check:* Replay protection is provided by using a monotonically increasing counter value at both ends or using a time stamp in the message. We have provided the counter value in a packet header that is used to defend replay attacks. A monotonically increasing counter of 32 bits is used at the both ends. But only the last 8 bits is sent in the packet for saving transmission and reception energy. The whole operation is given as follows

- Assume that both the receiver and sender are having the same counter value of 32 bits each.
- Operation of RC5-OCB is applied using a counter value of 32bits and MAC is obtained.
- While sending, full MAC (32 bits) is sent but we send only LSB 8 bits of counter value, thus saving transmission of 24 bits.
- Receiver calculates an expected counter (Cs) number depending on the last connection / synchronization.
- After receiving a packet the receiver will concatenate Cs (0-23) and received counter value (8bits). It will apply RC5-OCB operation to get MAC and cipher text and if the MAC is same, the packet is accepted, otherwise it will increment Cs and then try calculating MAC assuming some packet loss.

There should be some bound over increment and check approach given in step5, which can be done by setting a threshold value that depends upon the network packet loss rate. Also it can be minimized by application of bloom filter [16] .Chin et. al [28] provides a LOFT protocol that recommends sending of only 3bits in packet. It will not suit for broadcast communication, so we have taken a counter value of 8bits to make our scheme suitable for unicast as well as broadcast communication. To make this protocol more resilient to replay attacks, base station can broadcast the periodic counter value.

## V. SIMULATIONS AND ANALYSIS

For implementation of our proposed framework, we have extended Castalia simulator based on Omnet++ by adding ISA (Intelligent Security Agent).The approach proposed in this paper stresses on group communication and diversity of application scenarios, so we have tested it on three different application scenarios i.e. 'Military Surveillance System', 'Habitat Monitoring' and 'Agricultural Farming'. These three different application scenarios correspond to high, medium and low level security respectively. Table 1 provides the different security requirements of these scenarios. A comparison is done between using fixed security level and variable security level as shown in figure 3-5.

Military Surveillance System uses high security level (i.e. highest encryption level) most of the time, whereas in case of Agricultural Farming, only data authentication with low level of encryption is used.

Saving in energy is achieved by variable encryption level and flexibility of authentication and counter sending. ISA, depending on the current percept, will determine an adaptive reaction for level of security that would incorporate many policies, thus recommendations can also be given at deployment level itself or afterwards. Here percept information is collected from various layers using cross layer interactions and from the resource Manger. Percept Information may include following information

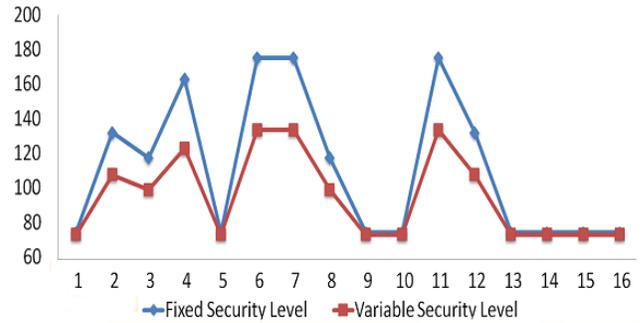

Figure 3

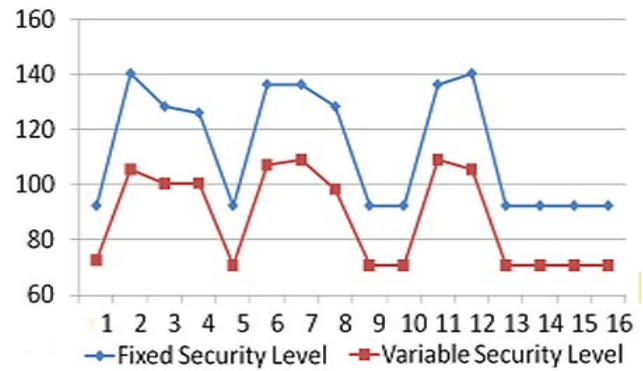

Figure 4

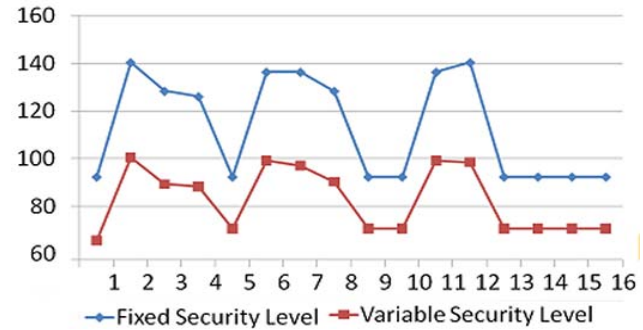

Figure 5

Figure 3, 4, 5. Military Surveillance System (Very Low Saving), Habitat Monitoring (Moderate Saving) and Agricultural Farming (High Saving). Y Axis -Energy Consumed (Joule) and X-Axis – Node IDs

Types of information considered for simulation are: available memory at that time, available energy, trust level of neighboring nodes, and predefined policies as well as recommendations.

The final conclusion that can be drawn from the result is that the function of ISA is prominent in case where the level of security to be achieved is known in advance like in case of 'Military application' the level of security is very high, thus the use of ISA is immaterial in this case as there will be very low energy saving whereas in case of 'Agricultural Farming' ISA plays an important role thus the amount of energy saved is very high. In case of "Habitat Monitoring" energy saved is moderate.

## VI. CONCLUSION AND FUTURE WORK

Improved security is especially important for the success of the wireless sensor network (WSN), because the data collected are often sensitive and the network is particularly vulnerable. While a number of approaches have been proposed to provide security solutions against various threats to the WSN, most of which are based on the layered design. We have pointed out that these layered approaches are often inadequate and inefficient. For our work to design a security scheme, we Considered one extra component viz. ISA (Intelligent Security Agent), which will interact with all the layers just like a resource manager and provides us with an extensive list of information.

Cross Layered approach is energy efficient and robust as shown by some of current research works. ISA helps in determining an adaptive reaction to security level. In our knowledge it is one of the first security frameworks that will provide security services to each layer and services of sensor networks. Through the simulation results, we have shown that energy efficient security could be achieved if we use variable security level for each application scenarios. We have simulated the above framework to test its feasibility, but the actual output will come from realistic implementation of this approach on sensor motes. So we are developing this framework using TinyOS as a security package. We have implemented a very raw form of ISA. Functions of intelligent security agent can be made more general and enhanced by employing efficient learning algorithm.

AUTHORS PROFILE

1. Kalpana Sharma: She's working as a Reader in the Deptt of Computer Science & Engineering in Sikkim Manipal Institute of Technology. She's done her M.Tech from IIT Kharagpur.

2. M.K. Ghose: Presently serving as the Head of CSE Deptt, Sikkim Manipal Institute of Technology, Sikkim , India. He's a Ph.D holder and has a number of publications in the field of Remote Sensing and GIS, Bioinformatics etc.

3. Kuldeep: He has completed his B-Tech from Sikkim Manipal Institute of Technology in 2009 June.